# TITLE PAGE

**Title:** Clinical transfusion-outcomes research: A practical guide

**Authors:** Sarah J Valk[1,2], Camila Caram-Deelder[2], Rolf. H.H. Groenwold[2], Johanna G van der Bom[2]

1. Jon J van Rood Center for Clinical Transfusion Research, Sanquin/LUMC, Leiden, The Netherlands,
2. Department of Clinical Epidemiology, Leiden University Medical Center, Leiden, The Netherlands,

**Corresponding author:**
Johanna G van der Bom
e-mail: j.g.vanderbom@lumc.nl
phone number: +31 71 526 5623
fax number: +31 71 526 6994

**Sources of support:**
This research was funded by Sanquin Research (grant PPOC-18-03).

**Manuscript counts:**

| | |
|---|---:|
| Manuscript | 3,481 words |
| Figures | 1 |
| Tables | 2 |
| References | 20 |

**COI statement:**
The authors declare that they have no conflict of interest relevant to the work presented in this manuscript.

**Running headline:**
Clinical transfusion-outcomes research

**Date:** 08/12/2025

**ABSTRACT**

Clinical transfusion-outcomes research faces unique methodological challenges compared with other areas of clinical research. These challenges arise because patients frequently receive multiple transfusions, each unit originates from a different donor, and the probability of receiving specific blood product characteristics is influenced by external, often uncontrollable, factors. These complexities complicate causal inference in observational studies of transfusion effectiveness and safety. This guide addresses key challenges in observational transfusion research, with a focus on time-varying exposure, time-varying confounding, and treatment-confounder feedback. Using the example of donor sex and pregnancy history in relation to recipient mortality, we illustrate the strengths and limitations of commonly used analytical approaches. We compare restriction-based analyses, time-varying Cox regression, and inverse probability weighted marginal structural models using a large observational dataset of male transfusion recipients. In the applied example, restriction and conventional time-varying approaches suggested an increased mortality risk associated with transfusion of red blood cells from ever-pregnant female donors compared with male-only donors (hazard ratio [HR] 1.22; 95% CI 1.05-1.42 and HR 1.21; 95% CI 1.04-1.41, respectively). In contrast, inverse probability of treatment and censoring weighted analyses, which account for treatment-confounder feedback, showed no evidence of an association (HR 1.01; 95% CI 0.85-1.20). These findings demonstrate how conventional methods can yield biased estimates when complex longitudinal structures are not adequately handled. We provide practical guidance on study design, target trial emulation, and the use of g-methods, including a reproducible tutorial and example dataset, to support valid causal inference in clinical transfusion research.

**MAIN TEXT**

**Introduction**

Clinical transfusion research aims to provide insight into the benefits and harms of transfusions. Randomized controlled trials (RCTs) are the gold standard for determining causal effects due to their ability to minimize bias through randomization and blinding. Yet, RCTs are not always feasible. Moreover, in transfusion research RCTs face unique complexities. For instance, while clear cutoffs of storage duration for comparing fresh vs. older red blood cell transfusions would be ideal, blood product availability hampers the achievability of such well-defined treatment comparisons.[1] Additionally, RCTs typically have shorter follow-up durations and smaller sample sizes to minimize participant burden and costs, and physicians may not agree to expose patients to a product perceived to be inferior. These constraints can limit both the causal interpretation of the findings and the generalizability of RCT findings to broader patient populations and longer-term outcomes. Provided observational studies are designed and executed rigorously, emulating an RCT such that potential bias is sufficiently mitigated, evidence from observational studies reliably complements the evidence based on RCTs.[2-4]

In this commentary, we shine the spotlight on methodological aspects of longitudinal observational data in clinical transfusion research. The goal of this commentary is to inform readers and researchers of such studies, to provide practical guidance and to encourage discussion about consequences of choices made in the design and analysis of observational studies of blood products. Specifically, we *(1)* discuss intricacies of observational data of blood product characteristics, *(2)* present an overview of methods used in studies of blood product characteristics, *(3)* discuss these methods, including considerations for designing and analyzing clinical transfusion studies of donor and product characteristics, and *(4)* provide a tutorial for the use of marginal structural models in investigating transfusion exposures.

1. **The challenges pertaining to blood product characteristics research**

In order to understand the impact of the complexities of transfusion research, we need to first introduce the key epidemiological concepts that relate to them. From the standpoint of modern causal inference, identifying the minimally sufficient adjustment set of covariates from a directed acyclic graph (DAG) is the starting point for any researcher wishing to estimate causal effects from observational data.[3] A causal DAG identifies which variables to adjust for, and which not, to be able to estimate a causal effect of the exposure of interest on the outcome. Drawing the DAG can be challenging, as transfusion exposure investigations are complex studies, involving longitudinal data, often including time-varying confounding and censoring of follow-up. In contrast to single timepoint interventions, or 'point treatments', transfusions are given over time and therefore conventional approaches to adjust for covariates might not be appropriate. When wrestling with the complexities of transfusion exposures, researchers can apply the target trial emulation framework. This framework seeks to emulate the design and analysis of a hypothetical RCT to estimate causal effects from observational data and has recently been endorsed by regulators.[5] The target trial emulation framework comprises several key components present in randomized controlled trials including eligibility criteria, treatment strategies, treatment assignment, follow-up start (time 0), follow-up end, the primary outcome, and the causal contrast. By defining these components and specifying their counterpart in the observational study, researchers can mitigate biases arising from time-varying exposures and confounders. Key assumptions that apply to causal inference from observational data derived from the potential outcomes framework are:

exchangeability, which assumes that the treated and untreated groups are comparable with respect to other covariates (i.e. 'no unmeasured confounding');

positivity, which requires that each individual has a non-zero probability of receiving any of the treatment levels given their observed covariates (i.e. 'no contraindications for treatment');

consistency, a concept which states that the potential outcome under the observed treatment assignment is equal to the observed outcome (i.e. 'no multiple versions of the treatment').

These assumptions are important in order to draw valid conclusions from observational data and must be carefully considered in transfusion exposure investigations; a more extensive coverage of this topic can be found elsewhere.[3]

Now that these concepts are introduced, there are several specific challenges which contribute to the difficulty of studying efficacy and safety in the clinical transfusion setting. First, because every transfusion is linked to a specific donor, there is a wide variation in the pool of available blood products. Depending on the research question, particular products might be very common or very rare, potentially leading to limited statistical power. Second, patients are frequently exposed to multiple transfusions. Although restrictive transfusion practices have become more common, on average patients in the Netherlands receive two transfusions per transfusion episode, with more transfusions given depending on the indication.[6] Summarizing these first two challenges: transfusions are a sequence of events and if these events are rare, their evaluation may not be possible using observational data. Third, external factors (e.g. calendar time, patient blood group and geographic region) influence the probability of receiving a unit with any of these different characteristics. Last, the existence of a possible bidirectional relationship between donor characteristics and patient outcomes is a recent insight that warrants increased scrutiny.[7] While there are similarities between clinical transfusion research and observational studies of rare exposures, the combination of the above mentioned challenges results in a unique situation that warrants careful consideration (Box 1).

> **Box 1. Similarities and differences between clinical transfusion research and observational studies of rare exposures**
>
> 1. Donor characteristics are distributed in the pool of blood products, depending on the characteristics these may be rare or common
>
> 2. Multiple transfusions lead to a mixture of exposures in one patient, requiring a method that can deal with intercurrent events
>
> 3. There are external factors influencing the probability of exposure to a certain product
>
> *Up to here, similar to all observational studies of rare exposures, with the addition that patients are not likely to adhere to initial assignment of exposure, and positivity violations becoming more likely for rare exposures*
>
> 4. Treatment-confounder feedback relating to product characteristics

Now, circling back to the key epidemiological concepts described earlier, it becomes clear that the combination of rare exposures and the need for sustained exposure over time may mean at least one of these assumptions may not hold. Positivity requires the treatment under investigation to happen in routine practice, and is unlikely to hold when exposure is rare. An important consideration here is whether the violation of this assumption is structural or not. We here consider the violation of positivity to not be structural, i.e. any lack of observations with the combination of covariates within the data is due to chance, and not because of a contra-indication for certain covariate combinations. For example, a woman experiencing bleeding during childbirth and requiring a single red blood cell product will have the same probability of receiving a unit of male donor blood as a male trauma patient. We continue this commentary with the notion that while positivity violations should be identified, structural non-positivity does not preclude the estimation of a causal effect here.

Continuing, there are also challenge pertaining to exchangeability, specifically, the concept of sequential ignorability. In brief, sequential ignorability is the notion that, given the observed history of past treatments, covariates, and outcomes, the treatment at each time point is conditionally independent of future outcomes, or in other words, effectively random. Why is sequential ignorability at stake here? To answer this question, we need to take a look at how to handle intercurrent events, or censoring.

The treatment effect a study sets out to quantify, or estimand, in transfusion exposure studies is commonly defined as initiating and adhering to the initial exposure assignment, that is, the characteristics of the first received transfusion, which can be seen as a 'per-protocol' analysis. The exposure of interest is then compared to a chosen reference category. However, in longitudinal studies intercurrent events need to be taken into account. As patients are exposed to multiple transfusions over time, they often do not solely receive the same exposure category throughout their follow-up. The question arises, what should be done with the follow-up from these 'cross-over' patients? In an RCT, they would be analyzed in the group to which they were originally assigned, in an 'intention-to-treat' analysis. However, doing so in the observational setting would dilute the effect estimate and possibly obscure relations the researcher might be interested in.

The answer to the aforementioned question was generally thought to be: to adjust for the time-varying cumulative number of transfusions by censoring the follow-up time of patients when they no longer adhere to their earlier exposure category. Because the number of transfusions is associated with the exposure (a particular product characteristic), and the outcome (mortality), the causal effect of exposure to the product characteristic of interest is estimated by adjusting for the cumulative number of transfusions received over time. Follow-up should be included using time-varying approaches, because selecting only patients who continued to adhere to their initial transfusion exposure will lead to bias.[3] Thus, rather than standard adjustment for covariates at baseline, control for confounding when time-varying confounding is present requires adjustment for time-varying covariates during follow-up of individual patients and censoring of follow-up at the time of non-adherence to the initial transfusion exposure category. However, depending on assumptions about the reasons for non-adherence to the initial transfusion exposure category, more advanced statistical modelling techniques may be required. This is because, when non-adherence is both 1. affected by prior exposure and 2. informative of the outcome, traditional methods can fail, and consequently yield biased results. The described phenomenon is known as treatment-confounder feedback, which is discussed in more detail in the next section.

**2. Treatment-confounder feedback in studies of transfusion exposures**

When time-varying confounders are affected by prior treatment, traditional methods (e.g. stratification, matching, outcome regression) are generally not suitable for confounding adjustment, as these may adjust away part of the effect of the exposure, yet also introduce a spurious association between exposure and outcome by conditioning on a collider in the DAG, interfering with the causal path.[3] In studying any exposures that are tied to the subsequent probability of receiving additional transfusions, i.e. exposures associated with consistent product hemoglobin increment differences, this hence becomes a problem that can no longer be solved easily. We refer to this as treatment-confounder feedback by product hemoglobin content. This concept, previously described by Zhao et al.[7], is illustrated in Figure 1.

In Figure 1, panel A shows the partial DAG for the investigation of donor characteristics and mortality. The number of transfusions received over time (L) is associated with the probability of receiving female donor-only units (A) and the underlying disease severity (U) and is therefore part of the minimally sufficient adjustment set. Adjustment for L is required to estimate the effect of A on mortality (Y); this can be done using g-methods (depicted in panel A as the arrow going towards A being removed) but traditional methods would also be appropriate here. With exposure to female donor-only units, however, comes a decreased hemoglobin 'dose' and therefore an increased need for additional transfusions (panel B). This can be illustrated by creating separate timepoints for treatment A and confounder L, thereby providing the complete DAG for this research question (panel C). This DAG shows that adjustment for L using traditional methods is not appropriate when the combined effect of $A_t$ (treatment at timepoint t) and $A_{t+1}$ (treatment at timepoint t+1) is of interest, as L is now located in the causal path of $A_t$ on Y, in addition to being a confounder for the effect of $A_{t+1}$ on Y. Alternative methods, such as g-methods (which include inverse probability of treatment weighting of marginal structural models, the parametric g-formula, and g-estimation of structural nested models[3]), are required here.

Specific situations where extra attention is expected to be warranted are the previously mentioned studies on donor sex, and pregnancy history of the donor. Also, storage duration of blood products can lead to smaller hemoglobin increments, and irradiation of red blood cell products would similarly require caution if chosen as exposure, both potentially influencing the time to next transfusion and the outcome. Note that this is a non-exhaustive list, and researchers are encouraged to think carefully if their research question necessitates the use of alternative methods which can be used to estimate treatment effects in the presence of treatment-confounder feedback.

## 3. Appropriateness of methods applied in clinical transfusion research of product characteristics

Several statistical analysis methods have been applied in the field of transfusion product characteristics research (*Table 1*).

*Restriction approaches* were employed, assessing the risk of exposure for groups of patients that were exposed to a single exposure type, without time-varying components.[8] This method could be classified as a per-protocol analysis, and is at risk of introducing bias, as the patients who only received one type of exposure throughout the follow-up period are inherently different from those who receive more transfusions, and are removed from the analysis because they 'crossed over'. Specific for the clinical transfusion field, an example would be the selection of male-donor only and female-donor only exposure in a 'unisex' recipient cohort (i.e. selecting from a set of patients followed over time the ones who received only transfusions of single-sex donor origin). Selection based on classification at the end of follow-up is not appropriate when time-varying confounders have been identified, as this can lead to biased estimates of risk for transfusion characteristics.

*Time-varying exposure and confounding adjustment* has also been applied, with the number of units received with a specific characteristic included in the model as a continuous variable.[9-11] A potential pitfall in applying this method is the inclusion of continuous variables without properly taking into account nonlinearity.[11,12] Stratified Cox proportional hazards regression models with time-dependent exposures have recently been applied in this field.[10] The time-varying approach is not appropriate if there is treatment-confounder feedback, as it can lead to biased effect estimates.

Other possible analysis strategies include *inverse probability of censoring weighting*, to account for patients in certain exposure categories being more likely to receive additional transfusions and no longer being compliant to the initial blood product exposure, and therefore having to be censored.[7,13]

## 4. Example dataset with applied methods illustrating that some approaches can lead to biased results

We applied the above-described methods to an example dataset to allow for a comparison of their performance in a semi-controlled setting. For this dataset, the study population consisted of male patients included in an earlier publication[9]. These male patients received transfusions in one of five included hospitals between 2005 and 2015. The complete exposure information was sourced from the Dutch municipality registration[14] to overcome the limitation of the original publication where 44% of the units donated by female donors had missing information about the pregnancy history. In Table S1, patient and blood product characteristics are described for this example dataset. Associations described in Table 2 apply to the patient population from the original, earlier publication and data were not altered or manipulated. This, opposed to the dataset for which the results are described in Table S2, which underwent an anonymization procedure that removed the empirical data, for the purpose of a publicly accessible tutorial.

In Table 2, the risk for exposure to ever-pregnant donor-only units compared to the reference group of male-only unit exposure is presented for the three methods described in Section 2 applied to an example dataset. The inverse probability of treatment- and censoring-weighted analysis, estimating the average treatment effect of exposure to donors with a positive pregnancy history on mortality, shows no association is present (hazard ratio 1.01, 95% confidence interval 0.85-1.20). In contrast, the application of the time-varying adjustment method and restriction method give an estimate that is further away from 1, which is likely because of treatment-confounder feedback by hemoglobin increment differences between the two compared blood product exposures.

Of note, non-collapsibility of the conditional HRs estimated using restriction and time-varying approaches results in an overestimation of the effect of exposure even in settings without treatment-confounder feedback. Depending on the distribution of covariates in the data that is being studied, this could further impact the performance of a chosen method.[15] It is also important to acknowledge that hazard ratios serve as valuable measures for assessing associations between variables in survival analysis; however, they do not directly measure causal effects. We therefore emphasize the distinction between individual-level and population-level interpretations of causal estimands. What's more, the choice of a cutoff of 10 for weight truncation, although commonly accepted in literature[16] to stabilize estimates and confidence intervals, could further impact the estimation of an effect, especially in situations where positivity violations occur. In conclusion, statistical choices have considerable influence on the conclusions that can be drawn from an investigation of blood product characteristics.

## 5. Tutorial for the application of marginal structural models as a way to estimate causal associations in the presence of treatment-confounder feedback

The use of inverse-probability weighted marginal structural models is not widespread in the field of clinical transfusion research, because their importance for studying transfusion exposures has not been recognized until recently. By providing an open-access example dataset with donor and patient characteristics, as well as concise R code, we hope to engage the scientific community, and encourage researchers to be more aware of the specific problems that arise when studying donor and product characteristics that relate to product hemoglobin content.

We provide a structured tutorial to perform the inverse probability of censoring weighting method described in Section 3 on a provided dataset (*Supplemental materials*, page 4). The dataset used in Section 4 is made available, after having applied an anonymization procedure to avoid sharing of personal patient data, and can be requested from the authors. The results for the inverse probability of censoring weighted analysis applied to the anonymized dataset can be found in Table S2. Because the original structure in the dataset was lost, all methods perform similar and can be interpreted to be unbiased due to the absence of treatment-confounder feedback. The dataset serves the purpose of applying the methods in practice and gaining insight into their implementation for the researchers' own work, and can be adapted to suit their needs, for example with regard to choices made on cutoffs for truncation.

## 6. Conclusions

The importance of thorough epidemiological study design in clinical transfusion research cannot be overstated. In this commentary, recent insights about hemoglobin increments and their impact on blood product characteristics research were extensively discussed, and an overview including an appraisal of these methods was provided. As an example, we made use of a large observational dataset of transfusion and patient data. We applied several methods used in the past and present, from which inverse probability of censoring weighting should be considered in the presence of treatment-confounder feedback because this method can adequately account for time-varying confounding in the presence of such feedback. We also provide a detailed tutorial to guide those pursuing similar research.

Evidently, clinical transfusion outcomes research using observational data can be complex. Specifically for blood product characteristics research, these challenges include the adjustment for time-varying confounders, the censoring of follow-up time when mixed exposure occurs, and treatment-confounder feedback by product hemoglobin content. Target trial emulation can be a useful tool to avoid both basic mistakes, and more complex analytical pitfalls.[4] Of note, assumptions and decisions about the analysis are best specified up front, to avoid the problems associated with 'researcher degrees of freedom'.[17] When the aforementioned challenges are appropriately handled, it is possible to draw causal conclusions from observational transfusion data.

We emphasize that, while there are certainly limitations to several study designs used in the past, there is always a tradeoff between bias and precision where in some cases, a simpler method might be preferable. This can include the choice of changing the exposure of interest to single timepoint exposures, as opposed to sustained exposure over time. Researchers can and should give sufficient attention to the strengths and limitations of their chosen approach, and sensitivity analyses can be employed to test the impact of assumptions on the robustness of the estimate.

To conclude, we addressed the appropriateness of specific statistical methods in the presence of treatment-confounder feedback in the clinical transfusion research field and have provided guidance for future research. The suitability of any method depends on assumptions about the underlying causal relations in the data, and careful consideration about this is needed to ensure interpretations are valid.

**Data availability statement**

The original data used in this article and an earlier publication is available for inspection upon request. An anonymized dataset which can be used to run the provided syntax on is available. Anonymization was performed by random permutation.[18] Note: the original data structure is not completely retained following anonymization, but more advanced anonymization methods that can retain the original data structure have not yet been developed for survival analysis.[19]

**Supplementary materials**

The Supplementary materials contain the tutorial with syntax for use in R (*Supplementary materials*). Additional tables with results for the provided, anonymized dataset available from the authors are reported in Table S2.

Supplement: Clinical transfusion-outcomes research: A practical guide

**TABLES**

**Table 1.** Overview of methods used to study blood product characteristics as exposure

| Methodology | Description of application in clinical transfusion research | Can handle treatment-confounder feedback | References |
|---|---|---|---|
| Traditional methods – restriction approach | Selection based on exposure classification at end of follow-up<br><br>Stratification, matching, outcome regression (including propensity score regression adjustment and matching) | No | Middelburg, Alshalani[8,20] |
| Traditional methods – time-varying approach | Exposure and confounder information modelled as time-varying variables<br><br>Cox proportional hazards model with time-varying treatment and confounders | No | Caram-Deelder, Edgren[9-11] |
| G-methods – inverse-probability of censoring weighting | Time-varying exposure and confounder information used for reweighing population to mitigate bias due to treatment-confounder feedback<br><br>Cloning, censoring, and inverse probability weighting, inverse probability-weighted marginal structural models | Yes | Zhao[7], Valk[14] |

**Table 2.** Results for different methods applied to the example dataset

| Analysis | No. of Deaths | No. of Recipients* | HR (95% CI) |
|---|---|---|---|
| Restriction method | | | |
| Male (reference) | 1,916 | 6,430 | 1 (reference) |
| Ever-pregnant female | 207 | 770 | 1.22 (1.05-1.42) |
| Time-varying exposure and confounding adjustment method | | | |
| Male (reference) | 1,916 | 10,901 | 1 (reference) |
| Ever-pregnant female | 207 | 1,494 | 1.21 (1.04-1.41) |
| Inverse probability of censoring weighting method | | | |
| Male (reference) | 1,916 | 10,901 | 1 (reference) |
| Ever-pregnant female | 207 | 1,494 | 1.01 (0.85-1.20) |

*Population included all male transfusion recipients that were identified in both datasets[9] with approx. 10% of patients not identified in the new dataset because of changes to the hospital administration records. HR hazard ratio; CI confidence interval

Supplement: Clinical transfusion-outcomes research: A practical guide

**FIGURES**

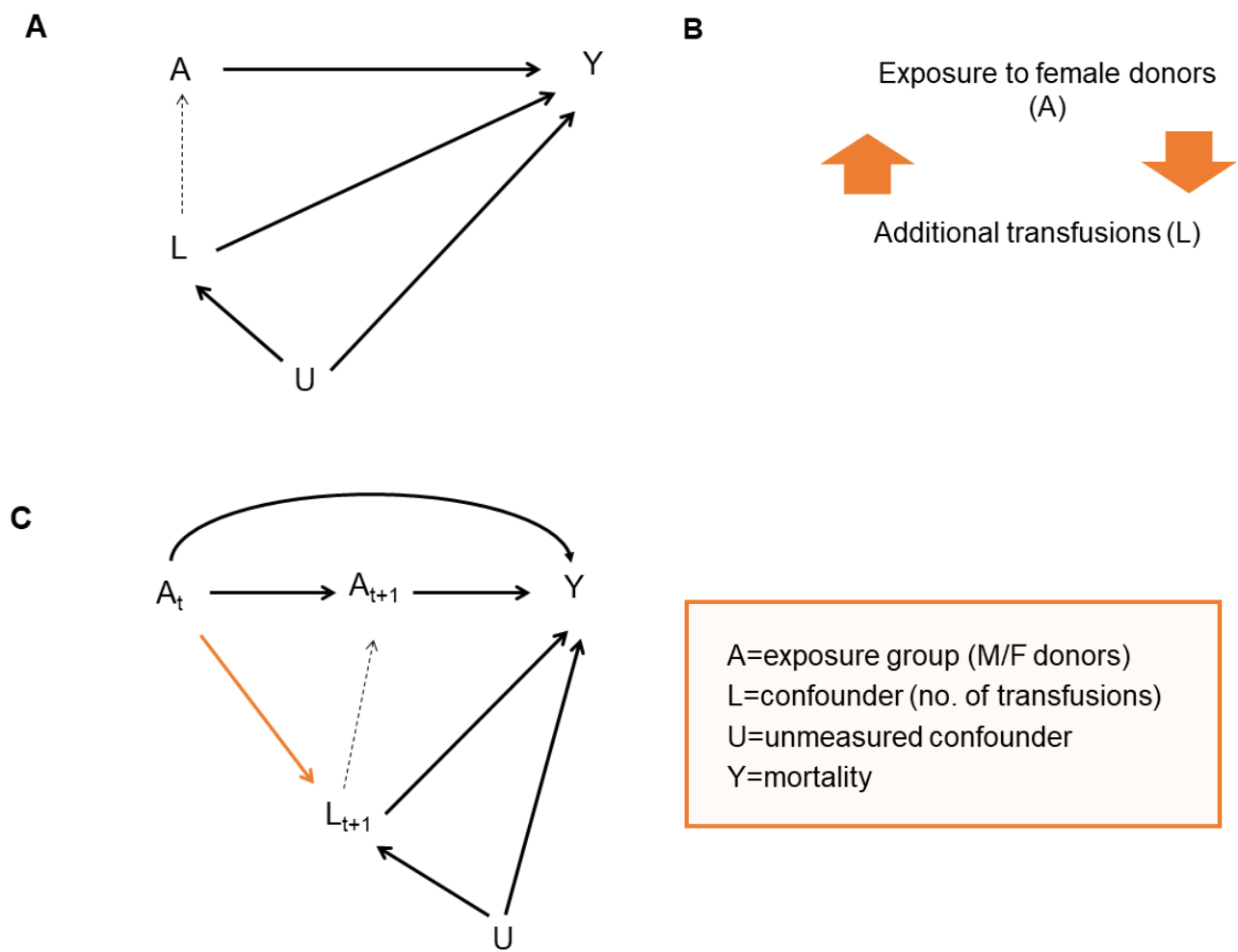

**Figure 1.** Different graphs to illustrate when advanced statistical modelling using g-methods is required.
**A)** Partial directed acyclic graph (DAG) of the effect of exposure to female donors (A) on mortality (Y) in transfusion recipients, confounded by unmeasured confounders (U, e.g. disease severity) through the cumulative number of transfusions (L). Dashed arrow represents the use of g-methods for the estimation of a causal effect of A on Y in the absence of treatment-confounder feedback, by removing the dependence of A on L.

**B)** Perceived bidirectionality if time is not taken into account, resulting in a cyclic graph, when assessing the effect of A on Y.

**C)** Complete DAG for the effect of exposure to female donor units including the treatment-confounder feedback over two timepoints (t, t+1) by lower hemoglobin concentration of units from female donors. Orange arrow represents the treatment-confounder feedback. Dashed arrow represents analysis using g-methods, removing the dependence of $A_{t+1}$ on L, making estimation of the causal effect of A on Y possible in the presence of treatment-confounder feedback.

# Supplemental material

# Clinical transfusion-outcomes research: A practical guide

Supplement: Clinical transfusion-outcomes research: A practical guide

**Supplement:** Clinical transfusion-outcomes research: A practical guide

**Contents:**

Supplement: Clinical transfusion-outcomes research: A practical guide

**Supplemental methods**

**Inverse probability of censoring weighting method (IPW)**

The dataset was organized as longitudinal survival data (with *t_begin* representing start of follow-up and *t_end* representing the end of follow-up for each patient row), for use in the *ipw* and *survey* package in R.[1] Initial follow-up is ordered as daily intervals for the first 28 days, followed by 4-week intervals ("blocks"). Weighted Cox proportional hazards models were fitted to correct for censoring and confounding.[1] Analyses were performed in R (version 3.6.3) and R Studio (version 2022.02.0+443) software.

The following variables were included in the multinomial logistic regression to estimate the baseline inverse probability of treatment weights: year of first transfusion exposure (*Transfusion_Year_first,* continuous), patient blood group (*Patient_ABORh*, categorical), hospital (*Hospital*, categorical). The outcome variable for the logistic regression was the categorical variable *Arm* (taking 0 if exposure was to the reference of male donors, 1 if exposure was to ever-pregnant female donors, and 9 if exposure was to other/mixed products).

The cumulative number of transfusions was included as the only covariate in the model for the generation of inverse probability of censoring weights (*Arm_Total_cum*), as a time-varying continuous variable. The outcome for this model was the censoring variable (*Censored*). Because patients could contribute multiple transfusion episodes, robust standard errors were used for the computation of the confidence limits.[2] Only patients exposed to reference (male, *Arm* taking the value 0) donors or exposure (ever-pregnant female, *Arm* taking the value 1) donors were included in the estimation of censoring weights. Censoring weights were generated for the dataset weighted by the inverse probability of treatment weights generated earlier. Weights were plotted within strata of follow-up time to determine the distribution of the weights with *ipwplot*.

The resulting weights were multiplied to create the final weights. Truncation, or trimming, of the weights in case of extreme weights (e.g. >10) is optional. The spread of the weights was assessed by calculating the $0.5^{th}$ and $99.5^{th}$ percentiles of the weights.

If patients were censored or died in a block, they were interval-censored. The actual end of follow-up, the variable *t_end_new,* was then used to replace the block time *t_end* for use in the Cox proportional hazards model.

The weighted Cox proportional hazards model was specified with the exposure (*Arm*), the outcome (*Death*), the time variables (*t_begin*, *t_end*) and the final weights. Only uncensored lines (*Censored* = 0) were included in the model.

A detailed R code including all steps described above is available at the end of the Supplemental materials (p. 4).

**Time-varying exposure and confounding adjustment method**

Cox proportional hazards models were fitted, adjusted for: cumulative number of transfusions (restricted cubic spline with three knots); hospital (categorical); blood group (categorical); calendar year (categorical); age of the donor (cumulative number of transfusions from donors aged >50 years, continuous); interaction term for cumulative number of transfusions and hospital.[3] Exposure is included as a binary, categorical variable.

Supplement: Clinical transfusion-outcomes research: A practical guide

This method is expected to be biased if treatment-confounder feedback is present due to limitations of traditional regression analysis. Analyses were performed in Stata, version 16 (StataCorp. 2019. Stata Statistical Software: Release 16. College Station, TX: StataCorp LLC).

**Restriction method**

Similar to method described above, with the distinction that only patients who received transfusions from the same exposure category as the first, are included and Cox PH regression is performed without a time-varying component.

This method conditions on information from the future follow-up of the patients, and is also expected to lead to bias. Analyses were performed in Stata, version 16 (StataCorp. 2019. Stata Statistical Software: Release 16. College Station, TX: StataCorp LLC).

## Tutorial for use of IPW for transfusion-outcomes research in R

The below provided syntax can be used to perform an inverse probability of treatment- and censoring-weighted analysis[3] for blood product exposures related to hemoglobin increment raising capacity of the product, on a provided, anonymized dataset. Note that this dataset does not retain all original features of the real dataset, and the treatment-confounder feedback structure was lost due to the anonymization process. The provided anonymized dataset is a representative example of a dataset generated with random permutation of the variables *Arm* (exposure, assigned randomly from original distribution), *Hospital* (category for the hospital where the patient received the transfusion, assigned randomly in one of four categories from original distribution of six hospitals)*, Patient_ABORh* (category of the blood group ABO and Rhesus type, assigned randomly from original distribution) and *Transfusion_Year_first* (year of the first transfusion of the patient, i.e. year of patient's start follow up, assigned randomly from original distribution). All other variables were kept identical to the original dataset.

**Tutorial syntax in R:**

The tutorial is organized as follows:

Step 0. Specify working directory and prepare files
Step 1. Inverse probability of treatment weights (IPTW) estimation with multinomial logistic regression
Step 2. Inverse probability of censoring weights (IPCW) estimation with weighted Cox regression
Step 3. Multiplication of weights (IPTW*IPCW) to create final weights
Step 4. IPW-corrected Cox model

Supplement: Clinical transfusion-outcomes research: A practical guide

**Step 0**

```
#Tutorial Clinical transfusion-outcomes research: A practical guide

#required: file = "Datafile-clinicaltransfusion.Rdata" (download available from BiorXiv)

##################################################
#Tutorial
#Male patients only
#Comparison: Male (0) vs Ever-pregnant female (1)

#Variables in the dataset are:
#PIN: unique patient identifier.
#Arm: 0: control, patients whose first transfusion was donated by a male donor; 1: exposure, patients whose first transfusion was donated by  a female donor who had been pregnant;  9: patients whose first transfusion was donated by blood donated by any other than exposure and control, i.e. female without history of pregnancy or sex of the donor unknown, and/or mixed exposure on day 1).
#Transfusion_Year_first: year of the first transfusion of the patient, i.e. year of patient's start follow up).
#Patient_ABORh: patient blood group, category.
#Hospital: hospital name, category.
#Censored: censoring indicator (0 if patient received all transfusions from the same Arm group, 1 if patient no longer adhered to initial group assignment).
#Arm_Total_cum: cumulative number of transfusions, continuous.
#t_begin and  t_end: time variables, each line refers to a single time period (t_begin refers to the start of the follow up, as required by the ipw package; all t_begin lines are rescheduled having -1 as reference; the first 28 days of follow up are included as one line per day and after day 28 the lines refer to blocks of 28 days).
#t_end_new: time variable, adjusted from block size (only blocks of 28 days are allowed) to real end of follow-up (individual days are allowed, e.g. if the patient has died at day 30, t_end_new would be '30', while t_end would be '56').
#Death: indicator for event at time t_end.

##################################################
#step 0. specify working directory and prepare files

#install packages
#install.packages("ipw")
#install.packages("survival")
#install.packages("survey")
#install.packages("dplyr")

#load packages
library(ipw)

library(survival)

library(survey)
```

```r
library(dplyr)

#set working directory: complete the path with the local where the datafil
e "Datafile-clinicaltransfusion.Rdata") is located
setwd("C:\\dir")

#clear workspace
rm(list=ls())

#load files
load(file= "Datafile-clinicaltransfusion.Rdata")
```

**Step 1-4**

```r
####################################################

#step 1. inverse probability of treatment weights (IPTW) estimation with m
ultinomial logistic regression
confounder_weight <- ipwpoint(exposure = Arm, family = "multinomial", nume
rator = ~1,  denominator =~Transfusion_Year_first + Patient_ABORh + Hospit
al, data = data)
```

```
#OUTPUT

# weights:   6 (2 variable)
## initial   value 925365.525208
## iter   10 value 345710.424406
## iter   10 value 345710.424400
## iter   10 value 345710.424397
## final     value 345710.424397
## converged
## # weights:  42 (26 variable)
## initial   value 925365.525208
## iter   10 value 420905.812229
## iter   20 value 388504.584486
## iter   30 value 361899.264984
## iter   40 value 349123.738592
## iter   50 value 345305.718723
## iter   60 value 345284.712165
## iter   70 value 345284.435350
## final     value 345284.422337
## converged
```

```r
data$iptwlogweights <- confounder_weight$ipw.weights
summary(data$iptwlogweights)
```

```
#OUTPUT

##    Min. 1st Qu.  Median    Mean 3rd Qu.    Max.
##  0.6927  0.9952  1.0009  1.0000  1.0066  1.3027
```

```r
#selection of subset of exposed (ever-pregnant, F1: coded as 1) and refere
nce (male, M: coded as 0); excluding Unknown, F0 and mixed (coded as 9) af
ter estimation of iptwlogweights
```

```
data<-subset(data, Arm!=9)

##################################################

#step 2. inverse probability of censoring weights (IPCW) estimation with w
eighted Cox regression
#IPTCW is estimated in the population weighted by IPTW
ipcwcox <- ipwtm(
  exposure = Censored,
  family = "survival",
  numerator = ~ 1,
  denominator = ~ Arm_Total_cum ,
  id = PIN,
  tstart = t_begin,
  timevar = t_end,
  type = "first",
  data = data,
  weight = data$iptwlogweights)

data$ipcwcoxweights <- ipcwcox$ipw.weights
summary(data$ipcwcoxweights)
```

```
#OUTPUT

##      Min.   1st Qu.   Median     Mean   3rd Qu.      Max.
##    0.0039    0.8961   0.9788   1.0004    1.0004 2384.5971
```

```
#plot IPCW weights
ipwplot(weights = ipcwcox$ipw.weights, timevar = data$t_end,
        binwidth = 1, main = "Inverse probability of censoring weights" ,
xlim = c(0, 28))
```

```
#OUTPUT
```

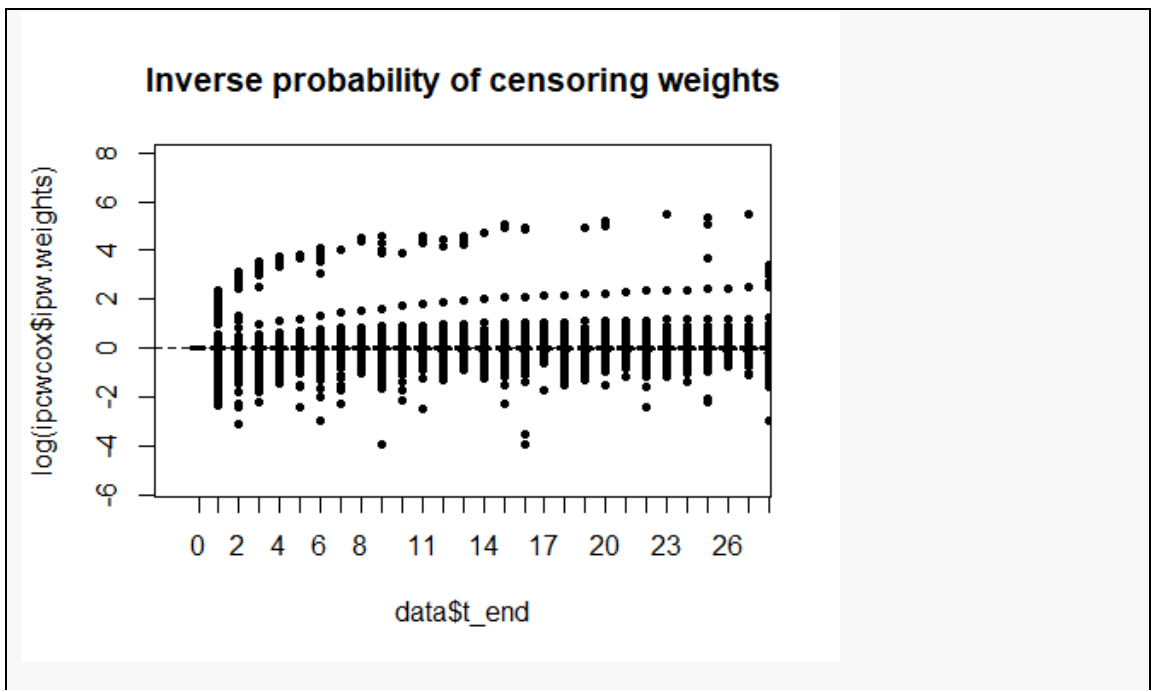

```
#interpretation:

#weights are depicted for the first 28 days; the distribution of the
weights is balanced with the exception of some large weights. Weight
s are selected for only the uncensored lines in step 3., leading to
less extreme weights.
```

```
#preparation of data for IPW-corrected model
#selection of non-censored observations only to limit the model to follow-
up time eligible for analysis (Arm=0 or Arm=1)
data2<-subset(data, Censored!=1)
#################################################

#step 3. multiplication of weights (IPTW*IPCW) to create final weights
data2$weights <- (data2$ipcwcoxweights*data2$iptwlogweights)
summary(data2$weights)
```

```
#OUTPUT

##    Min. 1st Qu.  Median    Mean 3rd Qu.    Max.
##  0.6753  0.8999  0.9664  0.9792  0.9995 60.2151
```

```
#store ranges of weights for assessment of extreme weights and weights dis
tribution
min <- min(data2$weights)
max <- max(data2$weights)
pct005 <- quantile(data2$weights, c(.005))
pct995 <- quantile(data2$weights, c(.995))

#store extreme weights
extreme <- subset(data2, weights>10)

#truncate weights (optional: large weights lead to instability of the IPW
estimator; truncation can reduce variance, but increase bias)
#data2["weights"][data2["weights"] >10] <- 10

#change t_end (to no longer be the 'block t_end', but the 'real t_end' fro
m patient final follow-up date)
data2$t_end <- data2$t_end_new

#################################################

#step 4. IPW-corrected Cox model
surveydesign1<-svydesign(id = ~ PIN, strata = ~ Arm, weights = ~ data2$wei
ghts,  data = data2)
summary(svycoxph(Surv(t_begin, t_end, Death) ~ as.factor(Arm), design = su
rveydesign1))
```

```
#OUTPUT

## Stratified 1 - level Cluster Sampling design (with replacement)
## With (12395) clusters.
## svydesign(id = ~PIN, strata = ~Arm, weights = ~data2$weights,
##     data = data2)
```

```
## Call:
## svycoxph(formula = Surv(t_begin, t_end, Death) ~ as.factor(Arm),
##     design = surveydesign1)
## 
##   n= 830334, number of events= 2297
## 
##                     coef exp(coef) se(coef) robust se     z Pr(>|z|)
## as.factor(Arm)1 0.01564   1.01576  0.06244   0.07283 0.215     0.83
## 
##                 exp(coef) exp(-coef) lower .95 upper .95
## as.factor(Arm)1     1.016     0.9845    0.8806     1.172
## 
## Concordance= 0.503  (se = 0.004 )
## Likelihood ratio test= NA  on 1 df,   p=NA
## Wald test            = 0.05  on 1 df,   p=0.8
## Score (logrank) test = NA   on 1 df,   p=NA
## 
##   (Note: the likelihood ratio and score tests assume independence of
## observations within a cluster, the Wald and robust score tests do not).
```

```r
msm <- svycoxph(Surv(t_begin, t_end, Death) ~ as.factor(Arm), design = sur
veydesign1)
a <- exp(coef(msm))
b <- exp(confint(msm))

#store counts for Deaths/Recipients, by exposure (0/1)
n_distinct(data$PIN)
```

```
## [1] 12395
```

```r
data00 <- subset(data2, Arm==0)
c <- n_distinct(data00$PIN) #Recipients 0, total
data01 <- subset(data2, Arm==0 & Death==1)
d <- n_distinct(data01$PIN) #Recipients 0, died

data10 <- subset(data2, Arm==1)
e <- n_distinct(data10$PIN) #Recipients 1, total
data11 <- subset(data2, Arm==1 & Death==1)
f <- n_distinct(data11$PIN) #Recipients 1, died

#create output table
Tutorialclinicaltransfusion <- data.frame(expcoef = a,
                                          confint = b,
                                          total0 = c,
                                          deaths0 = d,
                                          total1 = e,
                                          deaths1 = f,
                                          min = min,
                                          max = max,
                                          pct005 = pct005,
                                          pct995 = pct995,
                                          name = "Tutorialclinicaltransfusion")

#view output
View(Tutorialclinicaltransfusion)
```

Supplement: Clinical transfusion-outcomes research: A practical guideSupplement: Clinical transfusion-outcomes research: A practical guide

**Output**

| | expcoef | confint.2.5.. | confint.97.5.. | total0 | deaths0 | total1 | deaths1 | min | max | pct005 | pct995 | name |
|---|---|---|---|---|---|---|---|---|---|---|---|---|
| as.factor(Arm)1 | 1.01576 | 0.880632 | 1.171622 | 10901 | 1860 | 1494 | 263 | 0.6752993 | 60.21509 | 0.7735258 | 1.659287 | Tutorialclinicaltransfusion |

Labels:
- expcoef: HR
- confint.2.5..: 95% CI lower limit
- confint.97.5..: 95% CI upper limit
- total0: total in reference group
- deaths0: deaths in reference group
- total1: total in exposed group
- deaths1: deaths in exposed group
- min: minimum of final weights
- max: maximum of final weights
- pct005: 0.5th percentile of final weights
- pct995: 99.5th percentile of final weights

Supplement: Clinical transfusion-outcomes research: A practical guide

## Supplemental results

The characteristics of the anonymized dataset are presented in Table S1.

**Table S1.** Patient and product characteristics for the anonymized dataset

| Characteristics | Complete population | No-mixture subset* | Restriction subset[†] |
|---|---|---|---|
| Number of patients | N=18,206 | N=13,361 | N=7,659 |
| Number of deaths, (%) | 7,092 (39%) | 2,234 (17%) | 2,234 (29%) |
| Follow-up, median (IQR), days[‡] | 1,819 (389-2,744) | 341 (7-2,253) | 2,051 (679-2,977) |
| Person-time, sum in years | 87,382 | 42,999 | 41,107 |
| Age of patients, median (IQR), years | 65 (49-75) | 65 (44-75) | 64 (27-74) |
|     0 to 17 | 2,754 (15%) | 2,589 (19%) | 1,796 (23%) |
|     18 to 50 | 1,947 (11%) | 1,287 (10%) | 660 (9%) |
|     51 to 70 | 6,825 (37%) | 4,737 (35%) | 2,568 (34%) |
|     ≥71 | 6,680 (37%) | 4,748 (36%) | 2,635 (34%) |
| Transfusions of red blood cell units per patient, median (IQR) | 3 (2-6) | 2 (1-2) | 2 (1-2) |
| Units of red blood cells transfused, Number (%)[§] | 103,016 | 25,600 | 14,172 |
|     male donor | 65,239 (63%) | 22,454 (88%) | 12,617 (89%) |
|     female donor, ever-pregnant | 22,931 (22%) | 1,939 (8%) | 982 (7%) |
|     female donor, never-pregnant | 14,474 (14%) | 1,207 (5%) | 573 (4%) |

\* Consists of all the follow-up time during which patients either received all their red blood cell transfusions exclusively from one exposure category: female donors with a history of pregnancy (ever-pregnant donors), never-pregnant female donors, or male donors. The IPW analysis and Time-varying analysis use this definition. Follow-up time was censored at the time this inclusion criterion was violated.

† Consists of patients who received only one type of exposure (ever-pregnant, never-pregnant or male donor only) during the period in which they were followed up. Complete follow-up from these patients was included in the Restriction analysis.

‡ Median follow-up time is defined as the longest time any patient is in one of the comparisons. Exposure categories are: ever-pregnant donors and male donors.

§ Includes 372 (0.4%) transfusions with unknown donor sex and pregnancy history in the Complete population.

Below, the results for the anonymized dataset are presented (Table S2).

Supplement: Clinical transfusion-outcomes research: A practical guide

**Table S2.** Results for the different methods applied to the anonymized dataset

| Analysis | No. of Deaths | No. of Recipients | HR (95% CI) |
|---|---|---|---|
| Restriction method | | | |
|    Male (reference) | 1,860 | 6,316 | 1 (reference) |
|    Ever-pregnant female | 263 | 884 | 1.00 (0.88-1.14) |
| Time-varying exposure and confounding adjustment method | | | |
|    Male (reference) | 1,860 | 10,901 | 1 (reference) |
|    Ever-pregnant female | 263 | 1,494 | 1.01 (0.89-1.15) |
| Inverse probability of censoring weighting method | | | |
|    Male (reference) | 1,860 | 10,901 | 1 (reference) |
|    Ever-pregnant female | 263 | 1,494 | 1.02 (0.88-1.17) |

Here, due to the random permutation of the different variables, the original structure of the data was not maintained. Thus, the treatment-confounder feedback necessitating the use of the here described Inverse probability of censoring weighting method is not present, and all methods perform similarly. This, as opposed to the performance of these methods on the original data, as can be seen in Table 2 (p. 10) of the main article.